\documentclass[useAMS, usenatbib, usegraphicx]{mn2e}

\newlength{\colwidth}
\newcommand{\hMpc}{h^{-1}\,{\rm Mpc}}
\newcommand{\Msun}{{{\rm M}_\odot}}

\newcommand{\lya}{Ly$\alpha$}

\def\gtsima{$\; \buildrel > \over \sim \;$}
\def\ltsima{$\; \buildrel < \over \sim \;$}

\title[The 21~cm -- Galaxy Cross Power Spectrum]{LOFAR insights into the epoch of
  reionization from the cross power spectrum of 21~cm emission and galaxies}
\author[R.~P.~C. Wiersma et al.]
{R.~P.~C. Wiersma$^1$\thanks{E-mail:wiersma@mpa-garching.mpg.de}, 
B. Ciardi$^1$, R.~M. Thomas$^{2,3}$, G.~J.~A. Harker$^4$, S. Zaroubi$^5$, \and
G. Bernardi$^6$, M. Brentjens$^7$, A.~G. de Bruyn$^{5,7}$, S. Daiboo$^5$, V. Jelic$^7$, S. Kazemi$^5$, \and
L.~V.~E. Koopmans$^5$, P. Labropoulos$^{5,7}$, O. Martinez$^5$, G. Mellema$^8$,  A. Offringa$^5$, \and
V.~N. Pandey$^{5,7}$, J. Schaye$^9$, V. Veligatla$^5$, H. Vedantham$^5$, S. Yatawatta$^{4,7}$
\\
$^1$ Max Planck Institute for Astrophysics,
     Karl-Schwarzschild-Strasse 1, D-85748 Garching b. Muenchen, Germany\\
$^2$ Canadian Institute for Theoretical Astrophysics, M5S 3H8, Toronto, Canada\\
$^3$ Netherlands Institute for Neuroscience, Meibergdreef 47, 1105 BA, The Netherlands\\
$^4$ Center for Astrophysics and Space Astronomy, University of Colorado Boulder, CO 80309, USA\\
$^5$ Kapteyn Astronomical Institute, University of Groningen, PO Box 800, 
     9700 AV Groningen, the Netherlands\\
$^6$ Harvard Smithsonian Center for Astrophysics, 60 Garden Street, Cambridge, MA 02138, USA\\
$^7$ ASTRON, P.O. Box 2, 7990 AA Dwingeloo, the Netherlands\\
$^8$ Department of Astronomy and Oskar Klein Centre for Cosmoparticle Physics, AlbaNova, Stockholm University, SE-106 91 Stockholm, Sweden\\
$^9$ Leiden Observatory, Leiden University, PO Box 9513, 2300RA Leiden, the Netherlands\\
}

\begin{document}

\pagerange{\pageref{firstpage}--\pageref{lastpage}} \pubyear{2012}

\maketitle

\label{firstpage}

\begin{abstract}
Using a combination of N-body simulations, semi-analytic models and 
radiative transfer calculations, we have estimated the theoretical 
cross power spectrum between galaxies and the 21~cm emission from neutral hydrogen
during the epoch of reionization.  In accordance with previous studies, we find
that the 21~cm emission is initially correlated with halos on large scales (\gtsima 30~Mpc),
anti-correlated on intermediate ($\sim 5$~Mpc), and uncorrelated on small 
(\ltsima 3~Mpc) scales. This
picture quickly changes as reionization proceeds and the two fields
become anti-correlated on large scales. The normalization of the
cross power spectrum can be used to set constraints on the average neutral
fraction in the intergalactic medium and its shape can be  
a powerful tool to study the topology of reionization.
When we apply a drop-out
technique to select galaxies and add to the 21~cm signal the noise expected from the 
LOFAR telescope, we find that while the
normalization of the cross power spectrum remains a useful tool for
probing reionization, its shape becomes too noisy to be
informative.
On the other hand, for a Ly$\alpha$ Emitter (LAE) survey both the normalization
and the shape of the cross power spectrum are suitable probes of reionization.
A closer look at a specific planned LAE observing program using Subaru
Hyper-Suprime Cam reveals concerns about the strength of the 21~cm
signal at the planned redshifts. If the ionized
fraction at $z \sim 7$ is lower that the one estimated here, then using the cross 
power spectrum may
be a useful exercise given that at higher redshifts and neutral
fractions it is able to distinguish between two toy models with different topologies.

\end{abstract}


\section{Introduction}

The epoch of reionization (EoR) is considered one of the great
observational frontiers in astronomy today. It forms the crucial
bridge between the epoch of recombination and the galaxies we
currently observe. Since it was the era of first substantial galaxy
formation, it provides the context in which to understand the local
universe. The reionization process itself likely had a direct impact
on further galaxy formation and growth, primarily due to the dramatic
change in temperature of the Intergalactic Medium (IGM) caused by the
photo-ionization.

The investigation of the EoR takes place on both
theoretical and observational fronts. From the theoretical
perspective, analytic models of reionization have been employed since
\cite{Arons1972} and \cite{Hogan1979} and improved versions are still
being developed \cite[e.g.][]{Furlanetto2005,Choudhury.Ferrara_2006,BoltonHaehnelt07}.  
While these models
display a high degree of sophistication, the non-linear nature of
the feedback of reionization upon further galaxy formation is not typically
captured. Approaches based on (semi-)numerical methods can incorporate
such and other complexities, as for example the three-dimensional effects of shadowing and the overlap of ionized
regions. This allowed them to treat more of the physics properly, although this
comes of course at the cost of computing time.

All these theoretical
investigations have considerably advanced our understanding of the
progress of reionization, in particular 
the effect of inhomogeneities in the radiation field, the relative
importance of minihalos, quasars, and regular galaxies, and the
topology of reionization \cite[e.g.][]{Ciardi2000, Gnedin00, Razoumov2002,
  Ciardi.Ferrara.White_2003,Sokasian2003,Iliev2006, Kohler2007, Zahn2007,Thomas2011,
Ciardi.Bolton.Maselli.Graziani_2012}.

However, numerical simulations also have their limitations. To make
such calculations feasible for a representative sample of the Universe,
several assumptions/simplifications must be made. These
simplifications are a result of both the uncertainty in the physics
(e.g., star formation efficiency, properties of the ionizing sources,
escape fraction) and the finite computing resources which entail a certain
finite resolution for the simulation. Hence, to span the
parameter space of interest, methods typically resort to various Monte
Carlo and post-processing techniques that do not capture feedback
effects self-consistently \cite[e.g.][]{Ciardi2000, Thomas2011}.

On the observational front, probing the EoR directly has been
difficult and to date our main constraints on the time interval
during which reionization occurred are the Thomson scattering optical
depth at high redshift \citep{Komatsu2011} and the absence of
Gunn-Peterson troughs at the lower redshift end
(e.g. \citealt{Fan2006, Becker2007}, but see also
\citealt{Schroeder.Mesinger.Haiman_2012} for the detection of
Gunn-Peterson damping wings).

In pursuit of a direct detection of the EoR, a number of
instruments in various phases of development will be used to attempt to detect
neutral hydrogen via its 21~cm hyperfine transition.
PAPER\footnote{http://astro.berkeley.edu/\textasciitilde dbacker/eor/},
LOFAR\footnote{http://www.lofar.org/},
21CMA\footnote{http://21cma.bao.ac.cn/},
GMRT\footnote{http://gmrt.ncra.tifr.res.in/},
MWA\footnote{http://www.mwatelescope.org/}, and eventually
SKA\footnote{http://www.skatelescope.org} all hope to detect the 21~cm
signal from the EoR. It is not only the detection of the trend in the
decline (from higher to lower redshift) of the global neutral
hydrogen content that will be informative. Also of interest will
be the spatial distribution/fluctuations of the 21~cm signal at a
given redshift.

In these early attempts, detecting the 21~cm signal from the EoR will
be extremely challenging. To alleviate some of the problems, several
cross-correlation analyses with observations in other wavelength regimes
have been proposed. Under the assumption that the noise and
uncertainties will be mitigated by using two observations of such
different character, we can hope to put constraints on the nature
of reionization and hence gain further insights into the processes
active during 
the EoR. In recent years several authors have undertaken theoretical
studies of the cross-correlation analysis of 21~cm measurements with
other observations. Correlations with the cosmic microwave background
(CMB) \citep{Salvaterra.Ciardi.Ferrara.Baccigalupi_2005, Adshead2008,
Berndsen2010, Jelic_etal_2010}, galaxy surveys \citep{Lidz2009} and 
CO-emission surveys \citep{Lidz2011} have already been proposed.
This type of analysis is particularly timely also in view of the
exciting progress made in the observation of high-redshift galaxies
(e.g. \citealt{Ouchi2010,Bouwens_etal_2012} and references therein),
which is promising to provide a large, statistically significant
sample of such objects in the near future.

In this paper, we make predictions for the observation of the
cross-correlation between the 21~cm and galaxy fields along the lines
of \cite{Lidz2009}, but tailored towards the LOFAR-EoR experiment
and the future high-redshift
Subaru\footnote{http://www.naoj.org/} galaxy surveys. 
Using a dark
matter simulation and an efficient radiative transfer code, we begin
by cross-correlating the distribution of dark matter halos with the
distribution of the 21~cm signal. We continue by using a well-studied
semi-analytic model for galaxy formation and evolution to populate the
halos with galaxies, thereby incorporating realistic detection and
identification limits for the galaxies. We also add the expected noise
characteristics from LOFAR to the 21~cm signal to determine the use of
galaxy-21cm cross-correlation for detecting and characterizing
reionization.

This paper is organized as follows: \S \ref{sec-met} specifies the dark
matter simulation, radiative transfer code and the method used to
construct the cross power spectrum. In \S \ref{sec-the} we calculate
the cross power spectrum without imposing any observational limitations,
in order to find the theoretically predicted best possible scenario for
the detection. In \S \ref{sec-pred} we see how introducing more realistic
specifications for both the 21~cm and the galaxy survey modifies the 
theoretical result. Finally, in \S \ref{sec-con} we discuss these
results and the viability of performing such a cross-correlation in
practice.

\section{Method}
\label{sec-met}

Following \cite{Lidz2009}, we define the cross power spectrum between
the 21~cm emission and the galaxies as:

\begin{equation}
\begin{array}{lll}\Delta^2_{\rm 21,gal}(k) &=& \tilde{\Delta}^2_{\rm 21,gal}(k)/\delta T_{b0} \\
& = &  \langle x_{\textsc{hi}} \rangle \left[ \Delta^2_{x,{\rm
        gal}}(k) + \Delta^2_{\rho,{\rm gal}}(k) \right.\\
& & \left. + \Delta^2_{x\rho,{\rm gal}}(k)\right].
\end{array}
\label{eq-cps}
\end{equation}
The 21cm--galaxy cross power spectrum is thus made up from the sum of three
other cross power spectra, the neutral fraction--galaxy cross power spectrum,
$\Delta^2_{x,{\rm  gal}}$, the density-galaxy cross power spectrum, $\Delta^2_{\rho,{\rm gal}}$, and the neutral density--galaxy cross power spectrum, $\Delta^2_{x\rho,{\rm gal}}$. 
In Equation~\ref{eq-cps} we defined $\Delta^2_{\rm 21,gal}(k)$ such that the 21~cm
brightness temperature relative to the CMB for neutral gas at the mean
density of the universe $\delta T_{b0}$ is scaled out (normalized cross power spectrum), 
since the 21~cm field can be given by $\delta_{\rm 21}(r) = \delta T_{b0}
\langle x_{\textsc{hi}} \rangle (1 + \delta_x(r))(1 + \delta_\rho(r))$, where $\langle
x_{\textsc{hi}} \rangle$ is the mean volume averaged neutral fraction
and $\delta_i(r)$ represents the spatial field $i$ with respect to its
mean, e.g. $\delta_i(r) = (i(r) - \langle i \rangle)/\langle i
\rangle$. For the neutral hydrogen field, $i$ refers to $x_{\rm
HI}(r)$, the fraction of hydrogen that is neutral at position $r$, while for the
galaxy field, $i$ refers to $n_{\rm gal}(r)$, the number density of
galaxies at $r$. Finally, we work with the dimensionless cross power
spectrum, i.e. $\Delta^2_{a,b}(k) = k^3P_{a,b}(k)/(2\pi^2)$ for the
3D power spectrum and $\Delta^2_{a,b}(k) = 2k^2P_{a,b}(k)$ for the 2D
power spectrum, where $P_{a,b}$ is the dimensional cross power
spectrum between fields $a$ and $b$. We refer the reader to
\cite{Lidz2009} for a more detailed discussion of the three terms in
Equation~\ref{eq-cps}.

In order to construct the cross power spectrum, we therefore
require three fields, the density field, the neutral hydrogen field,
and the galaxy field.

For this work we make use of the well-studied {\it Millennium
  Simulation} \citep{Springel2005}. It is a dark matter simulation
featuring $2160^3$ particles in a $500 \hMpc$ comoving box run from $z
= 127$ down to $z = 0$. It was run in a $\Lambda$CDM cosmology with
$(\Omega_m, \Omega_\Lambda, \Omega_b h^2, h, \sigma_8, n) = (0.25,
0.75, 0.024, 0.73, 0.9, 1.)$, which implies a particle
mass of $1.2 \times 10^9 \Msun h^{-1}$. We have scaled the cosmology to the
more recent \textit{WMAP7} measurements found in (an early version of)
\cite{Komatsu2011}\footnote{The published version of the paper used an
updated version of the {\sc recfast} code and thus arrived at
\textit{slightly} different parameters.} -- $(\Omega_m, \Omega_\Lambda,
\Omega_b h^2, h, \sigma_8, n) = (0.272, 0.728, 0.02246, 0.702, 0.807,
0.961)$ -- in accordance with the method described in
\cite{Angulo2010}, scaling the output redshift, distance coordinates
and particle masses. All quantities are transferred to a $256^3$ grid,
using a cloud-in-cell scheme for the density and galaxy fields.

Halos with masses greater than $10^{10} \Msun$ (corresponding to a
limit of 20 particles) are selected as sources; not only do we reduce
resolution effects with such a cut, \cite{Lidz2009} used a similar
limit.    The spectrum assumed is that of a young, metal-poor
stellar population whose spectral energy distribution (SED) was
determined using {\sc starburst99} \citep{Leitherer1999} and is scaled
according to the mass of the halo. The
escape fraction of ionizing photons from each halo is taken to
be 10\%.  
The density and source field from
the simulation, together with the SED, are given as input to the {\sc
  bears} code \citep{Thomas2009} to calculate the neutral hydrogen
field. {\sc bears} is a radiative transfer code that, given the
luminosity of a source and its spectrum, calculates a spherically
averaged density profile around the source and embeds a spherically
symmetric ionization bubble. These bubbles are drawn from a catalogue
of 1D radiative transfer results of various types of spectra,
luminosities, redshifts and density profiles.  The code deals with
overlapping HII regions by increasing the sizes of the bubbles
involved in the overlap in such a way that the volume matches that of
the overlap regions, hence conserving photons. The reionization
histories calculated using {\sc bears} give a value of Thompson
scattering optical depth which is $\sim 0.09$ and within the 1-$\sigma$
error bar of the \textit{WMAP3} estimate.
For more details of the 1D radiative transfer code, the
implementation of {\sc bears} and its extensions, see
\cite{Thomas2008}, \cite{Thomas2009} and \cite{Thomas2011}.  

It has to
be noted that the resolution of the simulations does not allow us to
resolve the population of galaxies that are thought to be responsible
for the production of the majority of the ionizing photons during
reionization, which reside in halos that cool via atomic and molecular transitions
roughly in the range of $10^6 - 10^9 \mathrm{M_\odot}$ \cite[e.g.][]{Munoz2011,
Raicevic2011}. The clustering bias should, however, not be strong enough
to affect our results too much.

The output of the {\sc bears} code is the neutral fraction throughout
our simulation volume at different redshifts. To calculate the
21~cm-galaxy cross power spectrum we need to calculate the 21~cm
differential brightness temperature, which  is defined as follows \cite[e.g.][]{Thomas2009}:  
\begin{equation}
\begin{array}{ll}\delta T_b(r) = & 19 {\rm mK} (1 +
  \delta(r))\left(\displaystyle\frac{x_{\rm \textsc{hi}}(r)}{h}\right)
  \left(1 - \displaystyle\frac{T_{\rm CMB}}{T_s(r)}\right)\\
& \times \left[\displaystyle\frac{H(z)/(1+z)}{{\rm d}v_{||}/{\rm
        d}r_{||}}\right]\left(\displaystyle\frac{\Omega_b
    h^2}{0.02246}\right)\\
& \times \left[\left(\displaystyle\frac{1+z}{10}\right)\left(\displaystyle\frac{0.272}{\Omega_m}\right)\right]^{1/2},
\end{array}
\label{eq-Tb}
\end{equation}
where $\delta(r)$ is the matter overdensity at position $r$, $T_{\rm
CMB}$ is the CMB temperature, $T_s(r)$ is the spin temperature, and
the other symbols have their usual meanings. We make the
approximations that $T_s(r) \gg T_{\rm CMB}$ everywhere and that the
peculiar velocities do not contribute, such that the fourth and fifth
terms are unity \cite[see ][for a discussion on the contribution of
peculiar velocities to the 21~cm power spectrum]{Mao2011}.

{\bf When calculating cross power spectra and correlation functions (see
the following section) we bin the quantities such that $\Delta$log$k =
0.02$; for low $k$, we merge bins such that $\Delta k > 0.05$ to
ensure that our points are not correlated by the window function. For all figures in this paper, we plot quantities against co-moving
$k$.}

\begin{figure} 
\includegraphics[width=84mm]{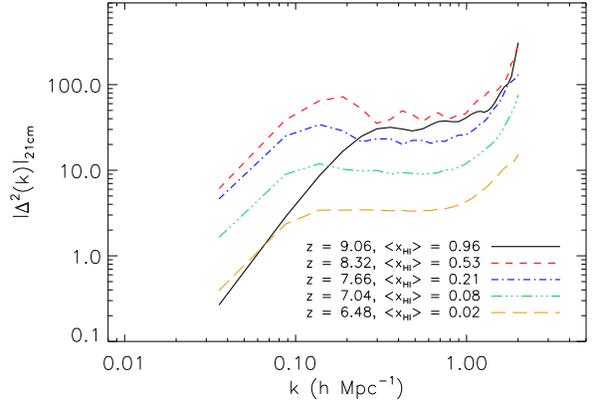}
\caption{The spherically averaged 3D 21~cm auto power spectrum for
  various redshifts/mean neutral fractions in our simulations.
  Power (particularly at small scales) decreases with decreasing
  neutral fraction.} 
\label{fig-theps}
\end{figure}

In Fig.~\ref{fig-theps} we show the spherically averaged 3D 21~cm
power spectrum for the redshifts we will concentrate on throughout
this work. The power spectrum shows the usual shape, 
with the normalization decreasing as the age of the universe (and mean ionized
fraction) increases, the large scale power being the last to
decrease. Initially there is much power on small scales due to the
clustered density profile. At later times, however, the small-scale
power diminishes as the large-scale power remains.

\section{The theoretical 21~cm -- halo cross power spectrum}
\label{sec-the}

Before concerning ourselves with what will be detected by upcoming
observations, it is useful to understand the intrinsic behaviour of
the 21~cm -- galaxy cross power spectrum. To study this, we use the
dark matter halos that we described in the previous section to
represent the galaxy field. Here we make no attempt to discern what
will be detectable or identified as a galaxy in the specified redshift
range.

\begin{figure} 
\includegraphics[width=84mm]{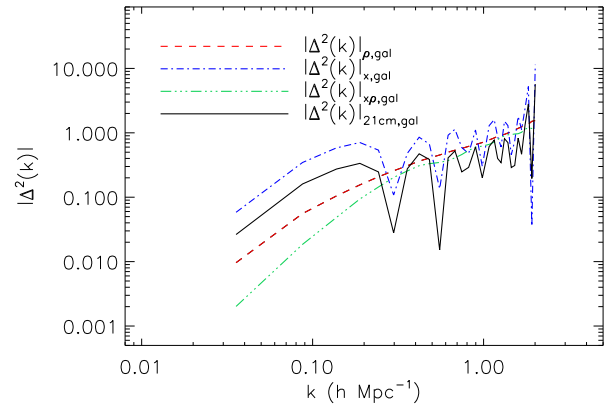}
\caption{The spherically averaged 3D 21~cm -- halo cross power
    spectrum for $z = 8.32$, corresponding to $\langle x_{\rm
    \textsc{hi}} \rangle= 0.53$. Also shown are the components of the
    21~cm -- halo cross power spectrum. }
\label{fig-the832}
\end{figure}

Fig.~\ref{fig-the832} shows the spherically averaged 3D 21~cm -- halo
cross power spectrum for $z = 8.32$, when the mean fraction of neutral
hydrogen is $\langle x_{\rm \textsc{hi}} \rangle = 0.53$. Here we have
broken it down into the three components listed in
Equation~\ref{eq-cps}, where the black solid curve is the final
result. It is worth remarking that, while $\Delta^2(k)_{\rho, {\rm
    gal}}$ is always positive, $\Delta^2(k)_{x \rho, {\rm gal}}$ is
always negative and $\Delta^2(k)_{x, {\rm gal}}$ is negative for $k$
where there are no oscillations. This implies an anti-correlation is
measured for the last two fields. 

Taking each term in turn, the density-galaxy cross power spectrum,
$\Delta^2(k)_{\rho, {\rm gal}}$, is the most straightforward
component to interpret. On small scales, these two fields correlate
very strongly, reminding us that halo formation is biased to high
density regions. The power decreases, however, towards large scales
at which halos are much less aware of the density. The increase of
power is roughly two orders of magnitude over two orders in magnitude
in scale.

The reionization process in our simulation proceeds in an `inside-out'
fashion, that is high density regions are typically ionized earlier
than low density regions. At the redshift we are considering here large-scale underdense regions are still
mostly neutral and free of galaxies, whereas the overdense
regions surrounding the galaxies have all become ionized. This
leads to an anti-correlation between the galaxy and neutral fraction fields, the
strength of which increases with decreasing scale, as illustrated by
the behaviour of the $\Delta^2(k)_{x, {\rm gal}}$ term in Fig.~\ref{fig-the832}.
Depending on the details of the model and the redshift, a turn around
manifests with a positive correlation on scales at which the galaxies
correlate with the density field.  On sub-bubble scales the
correlation is expected to die off because the interior of an HII
region is ionized independently from the local galaxy field.  The
typical size of the ionized bubbles is then imprinted on the cross
power spectrum at the smallest scales with an oscillatory
behaviour. This behaviour is less pronounced in \cite{Lidz2009}
because they can resolve smaller scales and lower mass halos. As a
result their reionization topology is less dominated by large
bubbles. Since bubbles are a generic feature of reionization, we do not
expect that the oscillations will disappear altogether in the high
resolution limit, but would tend towards the noise-like shape found in
\cite{Lidz2009}.

Finally, the $\Delta^2(k)_{x \rho, {\rm gal}}$ term, which is not
very significant on large scales, serves to cancel the galaxy -- density
cross power spectrum almost entirely at small scales (recall that they have
different signs). This effect is particularly strong for \textit{k}-modes
for which the size of the ionization bubble is imprinted. 

All three terms added give the measurable term, the 21~cm --
galaxy cross power spectrum. Its shape is mainly determined by the
$\Delta^2(k)_{\rho, {\rm gal}}$ and $\Delta^2(k)_{x, {\rm gal}}$ terms
on large scales, and almost completely by the latter at small
scales. This confirms the findings of \cite{Lidz2009} and we refer the
reader to their work for further discussion.

To better understand the behaviour of the cross power spectrum, we now
study two toy models. For the first model we begin with a neutral
universe and place ionized spheres with radius approximately $8 \hMpc$
around halos with mass greater than $10^{10} \Msun$ (this value gives
an ionized fraction of roughly $50 \%$). The second model is the dual
of the first: the background universe is taken to be ionized and the
bubbles contain neutral hydrogen. For both we use the same density
field and halo distribution as above, namely the one from redshift $z
= 8.32$, where  also the {\sc bears} simulation had $\langle x_{\rm
  \textsc{hi}} \rangle \approx 0.5$.

\begin{figure} 
\includegraphics[width=84mm]{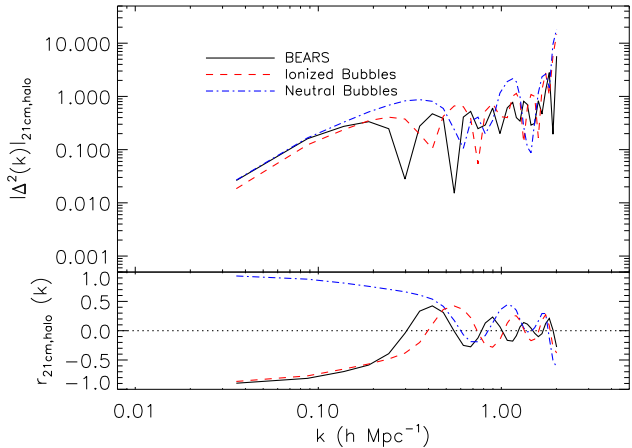}
\caption{The spherically averaged 3D 21~cm -- halo cross power spectrum
  (upper panel) and cross-correlation coefficient (lower panel) for
  two toy models at $z = 8.32$, $\langle x_{\rm \textsc{hi}} \rangle
  \approx 0.5$. The black solid line shows the result using {\sc
  bears}, while the red dashed line shows a toy model in which $8 \hMpc$ ionized
  bubbles are placed around halos, and the blue dot-dashed line shows
  a scenario where $8 \hMpc$ neutral bubbles are placed around halos
  in a fully ionized universe. In these extreme cases the general
  behaviour in the cross power spectrum is similar.  }
\label{fig-thetoy}
\end{figure}

We show the cross power spectrum
and the cross-correlation coefficient for these models, as well as the
prediction from {\sc bears} (black solid line in
Fig.~\ref{fig-the832}) in Fig.~\ref{fig-thetoy}. 
The first thing we notice is that
all three curves look remarkably similar. In the toy models the oscillations begin at
larger \textit{k} since the characteristic size of
neutral regions is larger compared to the radiative transfer
approach. The bottom panel shows the cross-correlation coefficient,
defined as $r(k) = P_{21, {\rm gal}}(k)/[P_{21}(k)P_{\rm
gal}(k)]^{1/2}$. Here we confirm that for the neutral bubble model the
21~cm emission is strongly correlated with the halo positions on large
scales. At smaller scales the halo density is uncorrelated
with the 21~cm emission since an $8 \hMpc$ bubble can contain many or
one halo.

Naturally, the size of our toy bubbles is arbitrary and has been chosen so that
the average neutral fraction is $\approx 0.5$ for all models. If we were to
use larger bubbles (they would need to be less ionized to maintain our
restriction of a global ionized fraction of roughly $50 \%$) we would
expect the oscillations to occur at slightly lower $k$, while we would
expect the opposite for smaller bubbles. This implies that the
characteristic size of the bubbles in the {\sc bears} simulation is
slightly larger than $8 \hMpc$, likely owing to the dominance of high
mass sources.

\begin{figure} 
\includegraphics[width=84mm]{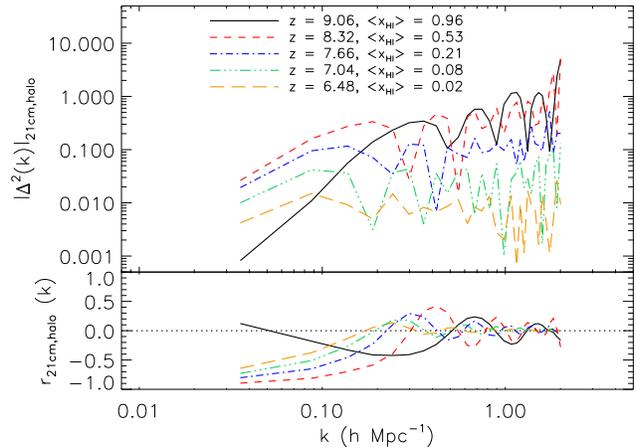}
\caption{The spherically averaged 3D 21~cm -- halo cross power spectrum
  (upper panel) and correlation coefficient (lower panel) for various
  redshifts/mean neutral fractions in our simulations. The
  anti-correlation is very strong on medium scales, but diminishes as
  the universe becomes more neutral.  }
\label{fig-allzsthe}
\end{figure}

Just as the normalization of the 21~cm auto power spectrum decreases with
decreasing neutral fraction, so does the 21~cm -- halo cross power
spectrum. This is seen in Fig.~\ref{fig-allzsthe}, where we plot the
cross power spectrum from the {\sc bears} results for our selected redshifts. Note that the scale
on which the oscillations occur increases with decreasing redshift as the
bubbles grow. The power spectrum also becomes flatter as the
correlations on small scales diminish.

The lower panel of Fig.~\ref{fig-allzsthe} shows the correlation
coefficient for the same set of redshifts. Here we see that at early
times ($z \sim 9$), the signals are strongly anti-correlated on intermediate scales
($k \sim 0.3-0.4$). This
is because the bubbles are still small. As the ionized regions become
larger, the signals become anti-correlated only on large scales.

All of this conforms to the results found in \cite{Lidz2009}. We
now turn to a more careful prediction of the observed signal.

\section{Predictions for the observed 21~cm -- galaxy cross power spectrum}
\label{sec-pred}
The spherically averaged 21~cm -- galaxy (halo) cross power spectrum
will not be directly measured by observations. In practice, the cross
power spectrum will be circularly averaged after the galaxy field has been
projected onto two dimensions, the galaxy sample will be constrained
by some selection criteria, and there will be substantial instrumental
effects for both the 21~cm signal and the galaxy sample.

We make use of the \cite{DeLucia2006} semi-analytic models (SAM) to generate
the necessary galaxy data. This is a well studied model that
roughly reproduces many $z = 0$ observations.  The semi-analytic
galaxy catalogue contains both rest-frame and observer-frame
magnitudes of many bands. To account for the change in cosmology, we
scale the magnitudes of the galaxies with the same mass factor
mentioned in section~\ref{sec-met}, and convert the wavelength of the bands
to account for the shift in redshift\footnote{Some of the calculations
in this work made use of the tool developed by \cite{Wright2006}.}. In
Table~\ref{tab-bands} we compare the original bands to the converted
bands. We caution that this model is designed to fit local universe
data, and that its predictions for the high-redshift universe may not be
correct. On the other hand, the SAM acts as a `best guess' since
fully analytic calculations would neglect the interplay between
galaxies, while hydrodynamic simulations would likely underestimate the
star formation rate at these redshifts \cite[e.g.][]{Schaye2010}.

To perform our predictions, we consider a very large observational
survey. A particularly ambitious program would cover {\bf 3 x 3 square degrees}
and overlap with previously well-studied fields (e.g. the Subaru Deep Field --
\citealt{Kashikawa2004} and GOODS -- \citealt{Giavalisco2004}).  We
consider surveys for two types of sources, high-redshift Lyman break galaxies
(LBGs) and Lyman alpha emitters (LAEs). The former uses broad band
photometry combined with a drop-out technique to detect the rest-frame 912 \r{A}
break. 
Such surveys typically use a strong colour cut to distinguish
LBGs from other objects \cite[e.g.][]{Bouwens2008}. They have the advantage
that existing filters can be used and combined with
already well-studied fields. One large disadvantage is that precise
redshifts cannot be determined without spectroscopic
follow-up.

LAEs, on the other hand, are objects that emit very strongly in the
1216 \r{A} \lya~line. They are typically found using narrow-band
surveys \cite[e.g.][]{Ouchi2010}. Since a narrow filter is used,
the redshifts of the objects are rather tightly constrained. The precise
nature of LAEs is, however, currently unknown. 

\subsection{Predictions for drop-out surveys}

We search for LBGs at all redshifts of interest, using the detection
limits of the survey presented in \cite{Ouchi2009a}. Their selection criteria
were that an object must not be detected in their 4 blue continuum
bands (${U, B, V, R} > {27.4, 28.0, 26.7, 27.0}$) and at the same time have a $z' - y$ colour greater than 1.5. We relax this
colour criterion to 1, and instead of looking at the \lya~trough as
they did, we consider the Lyman break since the positioning of our
observing bands is determined by the SAM and we can therefore not centre
so precisely on the \lya~trough.  
Table~\ref{tab-bands} shows how the SAM output bands are
affected by the redshift scaling and how they relate to the bands used
for the \cite{Ouchi2009a} selection.

Table \ref{tab-num} gives the detection efficiency of our galaxies for
each redshift of interest. Galaxies are most efficiently detected at
redshift 7.66 ($\langle x_{\rm \textsc{hi}} \rangle = 0.21$), likely
due to a combination of two factors.  First, a greater fraction of
galaxies are detectable than at higher redshift since the instruments
can detect galaxies with a lower absolute magnitude, and second, there
is a higher fraction of bright, star-forming galaxies than at lower
redshift. The total number of detected galaxies decreases at $z =
7.04$ ($\langle x_{\rm \textsc{hi}} \rangle= 0.081$) because for this
redshift the Lyman break falls in the middle of one of the bands, so
the colour difference cannot be efficiently detected.

\begin{table}
\caption{Central wavelength of bands in the observer frame of
 reference from the original SAM catalogue (left column), the
 catalogue converted to account for a different cosmology (central)
 and observed bands (right). See text for details.}
\label{tab-bands}
{\centering
\begin{tabular}{ccc}
\hline\hline Original SAM band & Converted band$^1$ & Corresponding \\
 & & observed band\\ 
\hline 
\textit{u} (3500 \r{A}) & 2947 \r{A} & \textit{U} (3600 \r{A}) \\ 
\textit{g} (4800 \r{A}) & 4042 \r{A} & \textit{B} (4400 \r{A}) \\ 
\textit{r} (6250 \r{A}) & 5263 \r{A} & \textit{V} (5500 \r{A}) \\ 
\textit{i} (7700 \r{A}) & 6485 \r{A} & \textit{R} (6400 \r{A}) \\ 
\textit{z} (9100 \r{A}) & 7662 \r{A} & \textit{z'} (9100 \r{A}) \\ 
\textit{J} (12600 \r{A}) & 10610 \r{A} & \textit{y} (9860 \r{A}) \\ 
\hline
\end{tabular}}
\\
$^1$ The actual value varies slightly ($\pm$10 \r{A}) with redshift so we
give approximate values here.
\end{table}

\begin{table}
\caption{Number of galaxies in the simulation box (second column), those
 selected as observable LBGs (third column), and those selected as LAEs
 (final column). See text for details.}
\label{tab-num}
\centering
\begin{tabular}{cccc}
\hline\hline Redshift & Total number & Selected & Selected \\
& of galaxies in the SAM & LBGs & LAEs \\
\hline 
9.06 & 72791 & 130 & 153\\
8.32 & 189904 & 351 & 581\\
7.66 & 420591 & 768 & 1940\\
7.04 & 820361 & 421 & 5517\\
6.48 & 1436450 & 1661 & 13109\\
\hline
\end{tabular}
\end{table} 

In a drop-out survey without any spectroscopic follow-up, there will be no
radial distance information and all of the objects will be projected
onto the same plane. The filter width used for these telescopes is
less than the width of our simulation box, so we choose to project
only a random slab of the box with a thickness that roughly
corresponds to 1000 \r{A}, which is the typical full-width at half
maximum of the response function that is used in photometric
surveys. The SAM outputs a galaxy catalogue gridded according to a
count-in-cell scheme. For each line of sight through the slab, we
calculate the mean number density weighted by a Gaussian function
($\sigma = 0.25 l$, where $l$ is the thickness of the slab) to
approximate the filter response function since each filter has a
different response function anyway \footnote{We neglect the
  possibility of galaxies being coincidental along the line of sight
  since we expect the angular resolution to be high enough to make
  this effect minimal.}. The 21-cm signal is an equally weighted
average across the entire slab.

\begin{figure} 
\includegraphics[width=84mm]{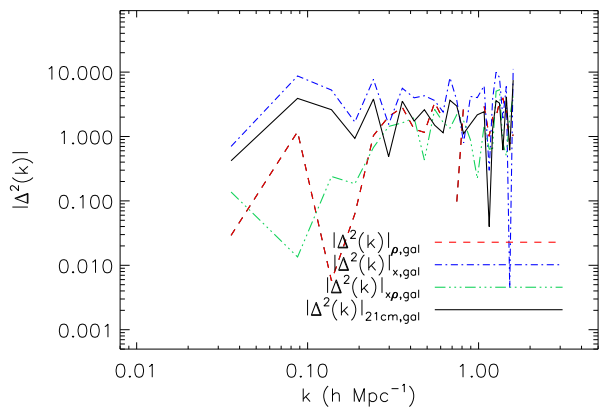}
\caption{The circularly averaged 2D 21~cm -- galaxy cross power
    spectrum for $z = 8.32$, $\langle x_{\rm \textsc{hi}} \rangle=
    0.53$ for a dropout survey. Also shown are the components of the
    21~cm -- galaxy cross power spectrum. The general shape and
    normalization of the power is recovered, but a number of features
    are lost.}
\label{fig-832obs}
\end{figure}

In Fig.~\ref{fig-832obs} we plot the projected, circularly averaged,
21~cm -- galaxy cross power spectrum predicted from our simulations
for $z = 8.32$ ($\langle x_{\rm \textsc{hi}} \rangle=
0.53$)\footnote{Before our results depended mostly on
the ionized fraction and the precise redshift was unimportant. The use
of the SAM has now made our predictions somewhat redshift
dependent.}. {\bf Note that since the galaxies are projected, $k$ describes
wavenumbers along angular distances.} We have again broken it into the individual components
from Equation~\ref{eq-cps}. Note that for this figure we have not yet
included the noise in the 21~cm signal.  We notice immediately that
the power spectra are much noisier, and the oscillations observed in
Fig.~\ref{fig-the832} are absent. On the other hand, some general
trends are retained, such as the increase of power with decreasing
scale. Also as before, the $\Delta^2_{x,{\rm gal}}(k)$ term defines
the shape for the 21~cm -- galaxy cross power spectrum. 

Next we add
noise appropriate to a 600~h observation with the LOFAR core to the
redshifted 21~cm signal. The actual 
LOFAR station positions are used to generate $uv$ tracks for a 4~h
observation at the zenith. From this we define a sampling function
$S(u,v)$ which describes how densely the interferometer baselines
sample Fourier space over the course of an observation, such that
$1/\sqrt{S}$ is proportional to the noise on the measurement of the
Fourier transform of the sky in each $uv$ cell. We generate
uncorrelated, complex Gaussian noise in each cell, but enforce the
condition $n(u,v)^*=n(-u,-v)$ where $n(u,v)$ is the noise in the cell
with coordinates $(u,v)$. This ensures that when we perform a
two-dimensional Fourier transform of this noise realization to obtain
a noise image, the image is real. The noise image is normalized to
ensure it has a temperature rms
\begin{equation}
\sigma_\mathrm{noise} = \frac{\lambda^2 T_\mathrm{sys}}{A_\mathrm{eff}\Omega_\mathrm{beam}\sqrt{2n_\mathrm{s}(n_\mathrm{s}-1) t_{int} \Delta\nu}},
\end{equation}
where $\lambda$ is the observed wavelength, $T_\mathrm{sys}$ is the
observed temperature, $A_\mathrm{eff}$ is the effective area of a
LOFAR station, $n_\mathrm{s}$ is the number of stations, 
$t_{int}$ is the integration time, $\Delta\nu$ is the bandwidth and
$\Omega_\mathrm{beam}$ is the area of the synthesized beam.  We take
$T_\mathrm{sys}=140+60(\nu/300\ \mathrm{MHz})^{-2.55}\ \mathrm{K}$ and
use values for the effective area tabulated on the ASTRON LOFAR
webpage\footnote{http://www.astron.nl/radio-observatory/astronomers/lofar-imaging-capabilities-sensitivity/lofar-imaging-capabilities-and-}.

Since some upcoming surveys are expected to cover very large fields
(upwards of {\bf three by three square degrees}, corresponding to $\sim 60$ proper Mpc at
$z=6.48$), we take 4 random slabs (for a square configuration these correspond
to $\sim 77$ proper Mpc) through the box and average them to make our
predictions less susceptible to sample variance.

\begin{figure} 
\includegraphics[width=84mm]{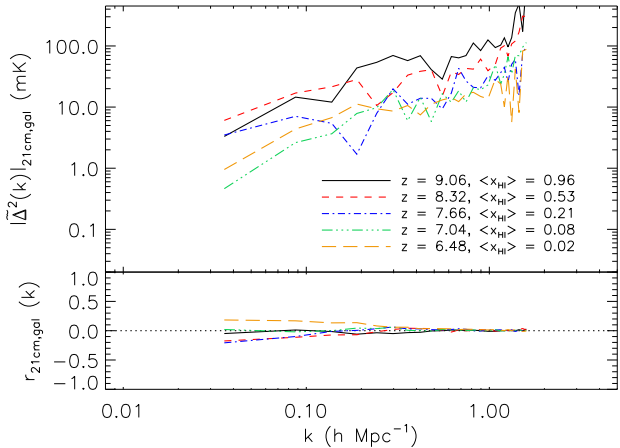}
\caption{The circularly averaged, unnormalized 2D 21~cm -- galaxy cross
  power spectrum (upper panel) and correlation coefficient (lower
  panel) for various redshifts/mean neutral fractions in our
  simulations in the dropout survey case. Although noisy, some of the
  general trends of the spherically averaged 3D power spectrum are
  recovered.}
\label{fig-allzsobs}
\end{figure}

In Fig.~\ref{fig-allzsobs} we present the result for our output redshifts
of interest. Please note that this figure shows the true measured (unnormalized)
cross power spectrum between the galaxy
field and the 21~cm emission, $\tilde{\Delta}^2_{\rm 21,gal}(k)$, as defined in
Equation~\ref{eq-cps}, and not the normalized one as in the previous figures. 
Very few of the trends seen in Fig.~\ref{fig-allzsthe} are recovered. The behaviour of increasing power with decreasing scale
is maintained, but the oscillations have all but disappeared. The
normalization of the cross power spectrum increases on all scales with
decreasing redshift, but the effect is small, such that it would be
difficult to distinguish between reionization states using the cross
power spectrum alone. The cross correlation coefficients (shown in the
bottom panel) are quite noisy and provide little information.

This indicates that using simple detection and selection techniques
will make it extremely difficult to glean information from the 21~cm
-- galaxy cross power spectrum. In our case, we simply do not detect
enough galaxies at the highest redshifts, too much information is lost
in the projection of the galaxy field, and the observing noise is too
large for a significant statement to be made.

\begin{figure} 
\includegraphics[width=84mm]{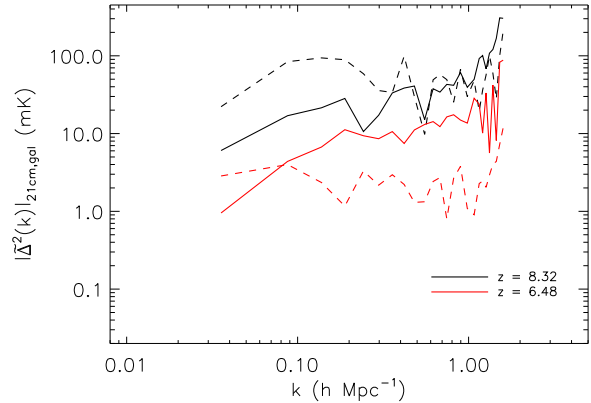}
\caption{The circularly averaged, unnormalized 2D 21~cm -- galaxy cross
  power spectrum calculated with (solid lines) and without (dashed
  lines) LOFAR noise in the dropout survey case. The noise inherent in
  the LOFAR instrument causes the two measurements to become similar.}
\label{fig-noiselbg}
\end{figure}

Fig.~\ref{fig-noiselbg} shows the impact that the LOFAR noise has at 
$z = 6.48$ (red lines)
and at $z = 8.32$ (black lines). Here we see that in the absence of noise
(dashed lines), the two redshifts appear to be quite distinct. 
When noise is introduced (solid lines), the curves appear more similar.
While at large scales adding the noise always decreases the
power spectrum, at small scales its effect depends on redshift.
In fact, unlike for the auto power spectrum where adding the noise always
adds to the power, in the cross power spectrum adding the noise moves
the spectrum towards what it would be in the case where the galaxy
field is crossed with a pure noise field. This means that at $z=8.23$
the cross power spectrum between the pure noise and galaxy fields
is lower than the one of the 21~cm and galaxy fields, whereas at $z=6.48$ the
21~cm signal is reduced by a much larger factor than the noise is,
resulting in opposite behaviour. This is seen only at small scales
because these are the ones where the drop in signal is more significant
(see Fig.~\ref{fig-allzsthe}).
In other words, after the noise is introduced, the $\Delta^2(k)_{x \rho, {\rm gal}}$ and
$\Delta^2(k)_{x, {\rm gal}}$ terms have diminished influence, while the
dominant term becomes $\Delta^2(k)_{\rho, {\rm gal}}$.

\begin{figure} 
\includegraphics[width=84mm]{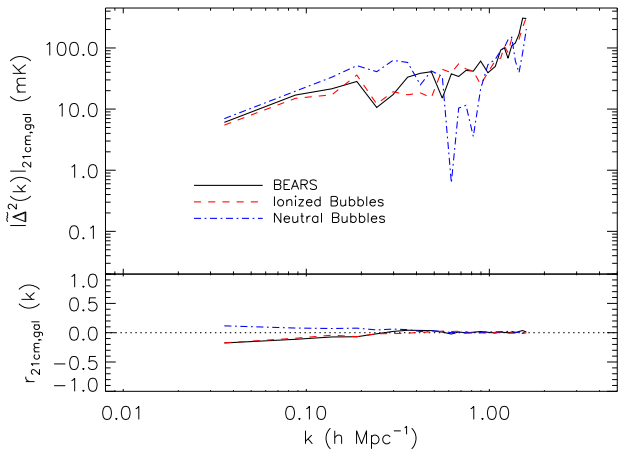}
\caption{The circularly averaged, unnormalized 2D 21~cm -- galaxy
  cross power spectrum (upper panel) and correlation coefficient
  (lower panel) for two toy models at $z = 8.32$, $\langle x_{\rm
  \textsc{hi}}\rangle = 0.53$ in the dropout survey case. The
  correlation coefficient is able to distinguish between the two
  scenarios.}
\label{fig-obstoy}
\end{figure}

Finally, we revisit our toy models to determine if the observed cross
power spectrum can tell us anything about the topology of
reionization. We have used the same toy models mentioned in the
previous section and have applied the same procedure described in this
section.  In Fig.~\ref{fig-obstoy} we see that these two very different
reionization scenarios result in a similar cross power spectrum. There
is certainly more power on medium to large scales in the neutral
bubble case, indicating that the cross power spectrum could help
identify different scenarios. One area in which these two scenarios
were distinct in section~\ref{sec-the} was the cross-correlation
coefficient. In the lower panel of Fig.~\ref{fig-thetoy} they were
markedly different on large scales. In Fig.~\ref{fig-obstoy}, the
difference is best noticed on medium scales.

\subsection{Predictions for LAE surveys}

\cite{Ouchi2010} report how they refined their dropout technique using
a narrow band filter to identify LAEs. They then obtained
follow-up spectra to establish the redshifts of their galaxies. While
only a narrow band in redshift is probed, the three dimensional positions 
of the objects can be established much more precisely.

We can therefore follow a similar procedure as we did in the previous
section, with a few slight modifications. The width of the narrow-band
filter used by \cite{Ouchi2010} is 132 \r{A}, which roughly
corresponds to the a slab about half the thickness considered in the
previous section. Since their filter is focused on a very narrow band that we
do not have access to in the semi-analytic model, we cannot select
based on the same colour criterion they used. Therefore, we use the
star-formation rate to intrinsic Ly$\alpha$ luminosity conversion
reported in \cite{Dayal2008}:
\begin{equation}
L_\alpha^{int} = 2.80 \times 10^{42} {\rm erg~s^{-1}} \displaystyle\frac{\rm
  SFR}{\Msun {\rm yr^{-1}}}.
\label{eq-laph}
\end{equation}

This will be attenuated and modified by a number of factors: the
escape fraction of ionizing photons, the fraction of Ly$\alpha$
photons that are destroyed by dust, gas inflows and outflows, and any
intergalactic absorption and scattering (see e.g.
\citealt{Dijkstra.Lidz.Wyithe_2007,Kobayashi.Totani.Nagashima_2007,
Dijkstra.Wyithe_2010,Jeeson-Daniel_etal_2012}).  The latter is expected
to be particularly relevant at these redshifts, when a substantial neutral
fraction is expected, but it is beyond the scope of this paper to investigate
this issue in more details. We defer further analysis to the future.
In order to select a
similar number of galaxies as \cite{Ouchi2010} (see below), we assume
the transmitted Ly$\alpha$ luminosity to be a factor of 150 less than
the intrinsic. 

\cite{Ouchi2010} report that they are sensitive to a Ly$\alpha$
luminosity of $2.5 \times 10^{42} {\rm erg s^{-1}}$ at $z = 6.56$.  We
have converted this using Equation \ref{eq-laph} to act as a
star-formation rate cut on our galaxies at each of our redshifts of
interest.  We use their detection limits to ensure that there is no
continuum emission bluewards of the Ly$\alpha$ break.

In Table~\ref{tab-num} we give the detection efficiency for the entire
box. Since we only take a slab, the actual number of galaxies
will be the number in the right-hand column adjusted for the
slab thickness.
We note that \cite{Ouchi2010} detected 207 LAEs at $z \approx 6.6$ in
a plane of similar size to ours; given that we use a slab thickness of
11 (out of 256) at this redshift, we only slightly overestimate the
number of detectable LAEs (albeit at a lower redshift).

\begin{figure} 
\includegraphics[width=84mm]{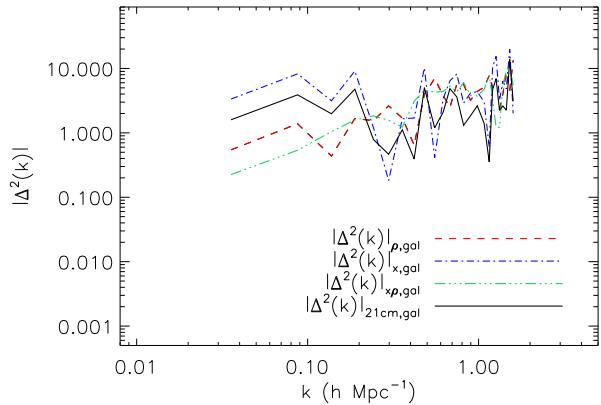}
\caption{The circularly averaged 2D 21~cm -- galaxy cross power
  spectrum considering a LAE survey for $z = 8.32$, $\langle x_{\rm
  \textsc{hi}} \rangle = 0.53$. Also shown are the components of the
  21~cm -- galaxy cross power spectrum. The general trend of
  Fig.~\protect\ref{fig-832obs} is maintained.}
\label{fig-832obs_LAE}
\end{figure}

Fig.~\ref{fig-832obs_LAE} is the analogue of
Fig.~\ref{fig-832obs}. Most of the features found in the dropout cross
power spectrum persist for the LAEs. Interestingly, the
$\Delta^2(k)_{x,~ {\rm gal}}$ term dominates even more here. Note that
for this figure we have not included the noise in the 21~cm signal.

\begin{figure} 
\includegraphics[width=84mm]{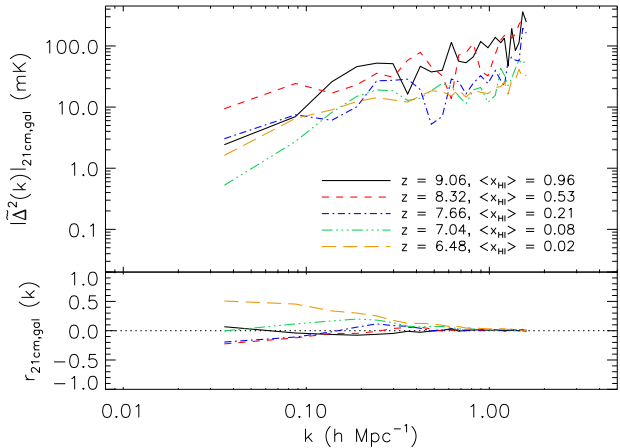}
\caption{The circularly averaged, unnormalized 2D 21~cm -- galaxy
 cross power spectrum (upper panel) and correlation coefficient (lower
 panel) for various redshifts/mean neutral fractions (see
 Fig.~\protect\ref{fig-allzsthe} for a legend) in our simulations in
 the LAE survey case. Slight differences in the cross power spectrum
 are seen for different redshifts.}
\label{fig-allzobs_LAE}
\end{figure}

In Fig.~\ref{fig-allzobs_LAE} we show the 21~cm -- LAE cross power
spectrum and correlation coefficient for a number of redshifts. Here
we have again averaged the calculation over 4 random slabs in the
simulation box and included the 21~cm noise. The result is similar to
that in the drop-out case (Fig.~\ref{fig-allzsobs}) in that many of
the features found in the theoretical case (Fig.~\ref{fig-allzsthe})
are not recovered. The cross-correlation coefficient, though, shows
more marked changes as reionization progresses. At high redshifts and
moderate ionized fractions, the correlation coefficient is negative,
but it turns positive for very low neutral fractions. The correlation
coefficient again appears to be the key to
drawing meaningful conclusions from these two measurements.

\begin{figure} 
\includegraphics[width=84mm]{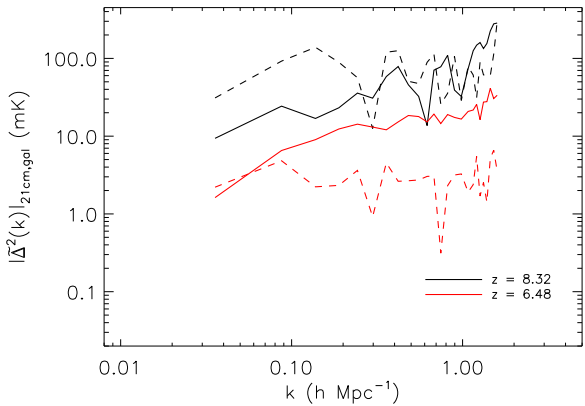}
\caption{The circularly averaged, unnormalized 2D 21~cm -- galaxy cross
  power spectrum calculated with (solid lines) and without (dashed
  lines) LOFAR noise in the LAE survey case. The addition of LOFAR
  noise makes it difficult to detect the suppression of power on small
  scales as the redshift decreases.}
\label{fig-noiselae}
\end{figure}

We briefly revisit the effect of noise from the LOFAR instrument on
our measurements. In Fig.~\ref{fig-noiselae} we repeat the
comparison we made in Fig.~\ref{fig-noiselbg}. Here we find that
at low redshift, while there is an intrinsic suppression of power on
small scales, the noise serves to mask that effect. On large scales
the noise decreases the power for both redshifts.

\begin{figure} 
\includegraphics[width=84mm]{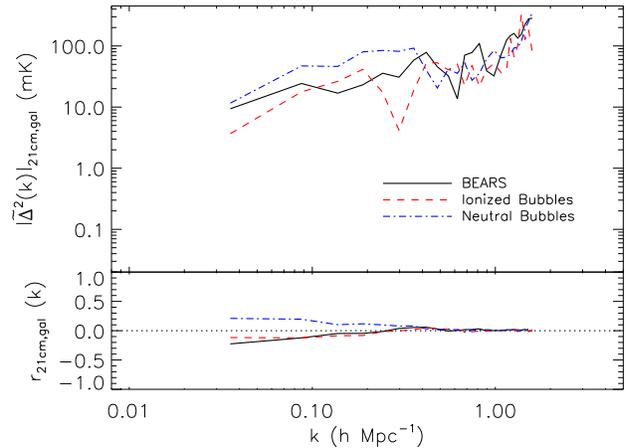}
\caption{The circularly averaged, unnormalized 2D 21~cm -- galaxy cross
  power spectrum (upper panel) and correlation coefficient (lower
  panel) for two toy models at $z = 8.32$, $\langle x_{\rm
  \textsc{hi}} \rangle = 0.53$ in the LAE survey case. It is possible
  to distinguish between the two ionization topologies using both the
  cross power spectrum and the correlation coefficient.}
\label{fig-obs_LAE_toy}
\end{figure}

A LAE survey could yield useful information about the progress of reionization,
but does it say anything about the topology of
reionization? In Fig.~\ref{fig-obs_LAE_toy} we return to our toy
models and apply the same calculation machinery. The result is a
somewhat clearer distinction between the two reionization
scenarios. At scales of roughly $5 \hMpc$, some difference is noted in
the cross power spectrum while at larger scales significant
differences in the correlation coefficient appear.

Finally, we consider four specific observing strategies for the Subaru
Hyper-Suprime Cam (kindly provided to us by Masami Ouchi). The details
are outlined in Table~\ref{tab-HSC}. Note that while there are {\it
deep} and {\it ultra-deep} observations planned for the $z = 6.6$
case, there is only an {\it ultra-deep} observation planned for the $z
= 7.3$ case. The current plan is to observe one of the $z = 6.6$ {\it
deep} fields with LOFAR, but here we consider all of the cases shown
in Table~\ref{tab-HSC}.

Our outputs do not line up precisely with the redshifts for these
planned observations, so we use the $z = 6.48$ and $z = 7.04$ outputs
and expect that the difference would be minimal. Furthermore, we use a
single slice of our simulation box to test the 3.5 and 4 square degree
cases (corresponding to $\sim 40$ proper Mpc at $z=6.48$) and average
9 slices for the 28 square degree case (corresponding to $\sim 107$
proper Mpc at $z=6.48$).  As for the estimates discussed earlier, we
convert the Ly$\alpha$ luminosities to equivalent star formation rates
using Equation~\ref{eq-laph}.

\begin{figure} 
\includegraphics[width=84mm]{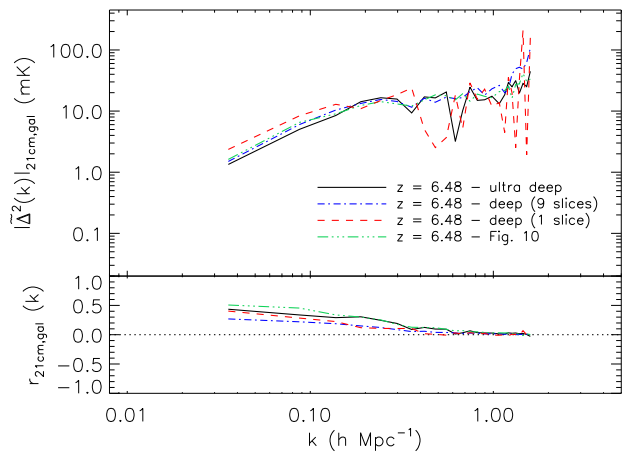}
\caption{The circularly averaged, unnormalized 2D 21~cm -- galaxy
  cross power spectrum (upper panel) and correlation coefficient
  (lower panel) for three specific Subaru Hyper-Suprime Cam observing
  strategies at $z = 6.48$. Also shown by the solid line is the $z =
  6.48$ analysis from Fig.~10. Little difference is seen between the
  four curves.}
\label{fig-obs_suprime6}
\end{figure}

\begin{figure} 
\includegraphics[width=84mm]{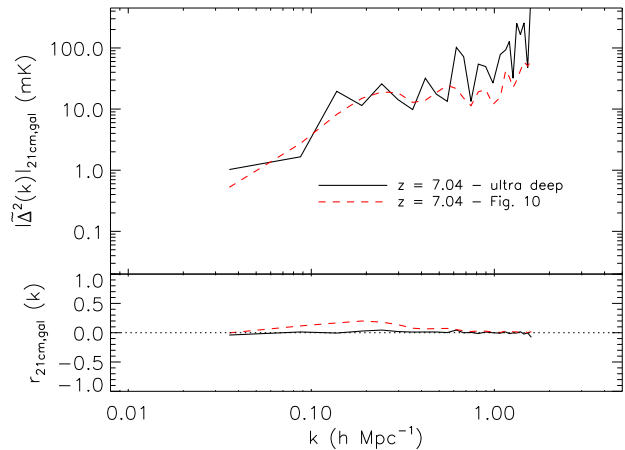}
\caption{The circularly averaged, unnormalized 2D 21~cm -- galaxy
  cross power spectrum (upper panel) and correlation coefficient
  (lower panel) for the Subaru Hyper-Suprime Cam observing at $z =
  7.04$. Also shown by the solid line is the $z = 7.04$ analysis from
  Fig.~10. Little difference is seen between the two curves.}
\label{fig-obs_suprime7}
\end{figure}

The results are shown in Figs.~\ref{fig-obs_suprime6} and
\ref{fig-obs_suprime7} for $z=6.48$ and $z=7.04$, respectively.  The
first noticeable thing is that it makes little difference when the
Ly$\alpha$ luminosity threshold or the observing area are changed by
factors of a few. As we saw in the previous section, a major component
in the shape of the curve is the LOFAR observing noise, not these two
factors. Averaging over more fields does seem to smooth the curves
slightly (although the curve in Fig.~\ref{fig-obs_suprime7} is
smoother only by chance), but it is not clear that using a deeper or
wider field will improve the measurement significantly. If our
fiducial reionization scenario is reasonable, these observations might
be better performed at higher redshift in order to obtain a stronger 21~cm signal.

\begin{table}
\caption{Characteristics of four observing strategies with the Subaru
   Hyper-Suprime Cam.} 
\label{tab-HSC}
\centering
\begin{tabular}{cccc}
\hline\hline Redshift & Number & Total area &  $L_{\alpha, {\rm min}}$\\
& of fields & (square degrees) & (erg/s) \\
\hline 
7.3 & 2 & 3.5 & $1.3 \times 10^{43}$ \\
6.6 & 2 & 3.5 & $2.8 \times 10^{42}$ \\
6.6 & 4 & 28 & $6.2 \times 10^{42}$ \\
6.6 & 1 & 4 & $6.2 \times 10^{42}$ \\
\hline
\end{tabular}
\end{table}

\section{Conclusions}
\label{sec-con}
We have endeavoured to make predictions for the 21~cm -- galaxy cross
power spectrum. Using our {\sc bears} code, we have performed
radiative transfer simulations on the well-studied \textit{Millennium
Simulation}. Beginning with the spherically averaged dimensionless
cross power spectrum between 21~cm emission and dark matter halos, we investigated
how these two fields relate before attempting to make predictions for
observations.

In general, we confirm the results of \cite{Lidz2009}, who made
their own predictions for the 21~cm -- halo cross power spectrum. We
find a similar shape and normalization, but owing to our coarser grid
and poorer resolution, we find that at large $k$ our power spectrum
shows oscillations related to the characteristic bubble size. 

The 21~cm emission is initially correlated with halos on large scales,
anti-correlated on medium, and uncorrelated on small scales. This
picture changes quickly as reionization proceeds and the two fields
become anti-correlated on large scales. Through toy models it becomes
apparent that these correlations can be an useful tool for inferring 
the topology of reionization. 

We then take the analysis in a different direction from
\cite{Lidz2009}. We attempt to make a more detailed mock observation
of this signal in order to make predictions for upcoming surveys,
using a well-studied semi-analytic model of galaxy formation and
evolution \citep{DeLucia2006} to bridge the gap between halos and
galaxies. Applying the drop-out technique used in the
\cite{Ouchi2009a} survey severely reduced the number of galaxies at
our disposal.

To further simulate the effect of observing and selecting real
galaxies, we considered only a slab of our simulation box
corresponding to the typical width of a filter. We then projected this
slice and circularly averaged the power spectrum. We also added the
noise expected from the LOFAR instrument to the 21~cm signal.

The result is that while the shape of the cross power spectrum is nominally preserved, 
its normalization seems to be the most
powerful tool for probing reionization. In particular, it is sensitive
to the ionized fraction as we show that different reionization
histories yield similar cross power spectra for the same ionized
fraction.

Compounding these problems is the fact that any galaxy survey will
likely focus on a specific drop out range. We have seen that the cross
power spectrum is quite useful when comparing the relative
differences between different models or redshifts. In the absence of
another redshift with which to compare results to, it might be
troublesome to come up with a robust statement about reionization.

We turned to a more precise measurement of the galaxy redshifts that
would be found using a LAE survey. We found that if the radial
position of the galaxy is known, then much more information about the
nature of reionization can be gleaned from cross-correlating the galaxy
and 21cm fields. Using a
LAE survey could in principle allow one to describe reionization using
both the shape and normalization of the cross power spectrum.

A closer look at a specific planned LAE observing program using the Subaru
Hyper-Suprime Cam reveals concerns about the strength of the 21~cm
signal at the planned redshifts. If our estimate of the ionized
fraction at $z = 7$ is too high, then using the cross power spectrum might
be a useful exercise given that at higher redshifts and neutral
fractions it is able to distinguish between toy models with two
different topologies. Indeed we predict that a detection of a
correlation signal will be made - the main issue will be the
interpretation of that signal.

There are a few observational effects which we have not included in
our analysis. {\bf We have neglected to include peculiar velocities
associated with both the galaxies and the IGM gas.} On the galaxy side, we have neglected to mimic the
effect of interlopers that could be misidentified as high-redshift
galaxies. On the 21~cm side, we have assumed that the projected signal
can be recovered on the angular scale of one of our resolution
elements. A greater source of uncertainty however, may be the effect
of foreground removal \cite[e.g.][]{Jelic_etal_2008, Bernardi_etal_2009, Harker_etal_2010, Jelic2010,
Petrovic2011}. 

The prospect of combining upcoming galaxy surveys with measurements
of the 21~cm signal remains quite exciting. Such combinations should
not only be able to tell us about the progress of reionization, but
the topology and the main drivers as well. This may turn out to be a
key exercise in understanding this epoch while true imaging of
21~cm maps remains pending.

\section*{Acknowledgments}
This work was supported by DFG Priority Program 1177.  The authors
would like to thank Masami Ouchi for kindly providing observing
strategies for the Subaru Hyper-Suprime Cam.  GH is a member of the
LUNAR consortium, which is funded by the NASA Lunar Science Institute
(via Cooperative Agreement NNA09DB30A) to investigate concepts for
astrophysical observatories on the Moon.
LVEK, HV and SD acknowledge the financial support from the European
Research Council under ERC-Starting Grant FIRSTLIGHT - 258942. {\bf We
would also like to thank the anonymous referee whose comments improved
this paper.}

\bibliographystyle{mn2e} 
\bibliography{ms}
\label{lastpage}

\end{document}